\documentclass[english]{sn-jnl}

\usepackage{graphicx}
\makeatletter
\let\saved@includegraphics\includegraphics
\AtBeginDocument{\let\includegraphics\saved@includegraphics}
\renewenvironment*{figure}{\@float{figure}}{\end@float}
\makeatother
\usepackage{multirow}
\usepackage{xcolor}
\usepackage{color}
\usepackage{amsmath}
\usepackage{bm}
\usepackage{babel}
\usepackage{float}
\usepackage{hyperref}
\hypersetup{
     colorlinks = true,
     citecolor  = blue,  
     linkcolor  = magenta 
}
\graphicspath{{figs/}}
\usepackage{lineno}

\jyear{2022}%

\theoremstyle{thmstyleone}%
\theoremstyle{thmstyletwo}%

\theoremstyle{thmstylethree}%

\begin{document}

\title{Non-Abelian gauge fields in circuit systems}

\author[1,2]{\fnm{Jiexiong} \sur{Wu}}
\equalcont{These authors contributed equally to this work.}
\author[1,2]{\fnm{Zhu} \sur{Wang}}
\equalcont{These authors contributed equally to this work.}
\author[1,2]{\fnm{Yuanchuan} \sur{Biao}}
\equalcont{These authors contributed equally to this work.}
\author[3]{\fnm{Fucong} \sur{Fei}}
\author[3]{\fnm{Shuai} \sur{Zhang}}
\author[2]{\fnm{Zepeng} \sur{Yin}}
\author[2]{\fnm{Yejian} \sur{Hu}}
\author[2]{\fnm{Ziyin} \sur{Song}}
\author[2]{\fnm{Tianyu} \sur{Wu}}
\author*[3]{\fnm{Fengqi} \sur{Song}}\email{songfengqi@nju.edu.cn}
\author*[1,2]{\fnm{Rui} \sur{Yu}}\email{yurui@whu.edu.cn}
\affil[1]{\orgname{Wuhan Institute of Quantum Technology}, \orgaddress{\city{Wuhan}, \postcode{430206}, \country{China}}}
\affil[2]{\orgdiv{School of Physics and Technology}, \orgname{Wuhan University}, \orgaddress{\city{Wuhan}, \postcode{430072}, \country{China}}}
\affil[3]{\orgdiv{National Laboratory of Solid State Microstructures, Collaborative Innovation Center of Advanced Microstructures, and School of Physics}, \orgname{Nanjing University}, \orgaddress{\city{Nanjing}, \postcode{210093}, \country{China}}}

\abstract{Circuits can provide a platform to study novel physics and have been used, for example, to explore various topological phases. Gauge fields -- and non-Abelian gauge fields, in particular -- can play a pivotal role in the design and modulation of novel physical states, but their circuit implementation has so far been limited. Here we show that non-Abelian gauge fields can be synthesized in circuits created from building blocks that consist of capacitors, inductors and resistors. With the building blocks, we create circuit designs for the spin–orbit interaction and the topological Chern state, which are phenomena that represent non-Abelian gauge fields in momentum space. We also use the approach to design non-reciprocal circuits that can be used to implement the non-Abelian Aharonov-Bohm effect in real space.}

\maketitle

Gauge fields are a key concept in modern physics, playing a role in high-energy physics, condensed matter physics, and electromagnetism. They can be categorized as Abelian and non-Abelian depending on whether their associated symmetry groups are commutative or not, with Abelian gauge fields generating distinct physics from non-Abelian ones. In recent years, non-Abelian gauge fields have been created in cold atoms\cite{Hofstadter_moth_2005,soc_coldatom_N2011,SGF_optical_2016,SGF_soc_S_panjianwei_2016,RevModPhys_ultracold_2019}, photonic systems\cite{Non_Abelian_prl_2014,SGF_nonAbelian_light_prl_2016,SGF_nonAbelian_light_nc_chenyuntian,optical_S_2019,SGF_nonAbelian_light_s_realspace,RevModPhys_photonics_2019,whittakeroptical2021}, polaritonic systems\cite{SGF_photons_Nc_2017}, and mechanical systems\cite{Fruchart2020}. Various novel physical effects related to non-Abelian physics have also been explored, including the quantum anomalous Hall effect\cite{cold_atom_Haldane_N_2014}, topological insulators\cite{RevModPhys_photonics_2019,RevModPhys_ultracold_2019}, non-Abelian monopoles\cite{Sugawa1429}, the real-space non-Abelian Aharonov-Bohm effect\cite{SGF_nonAbelian_light_nc_chenyuntian,SGF_nonAbelian_light_s_realspace}, non-Abelian band topology\cite{wu_non-abelian_2019}, and non-Abelian quantum simulation\cite{non_Abelian_PhysRevLett2015}.

A circuit that consists of lumped-parameter components is a system whose properties are determined by the parameters of the components and the topology of the electrical network. Since the wires are flexible and capable of making connections regardless of the spatial dimensions, the components can be connected with braided structures to achieve matrix-type tunnelling. As a result, circuits can be an ideal platform to study non-Abelian physics. Circuit systems have previously been used to implement various topological phases, including topological insulators and semi-metals\cite{CKT_TI_2015_PRX,CKT_TI_2015_PRL,CKT_2018_commPhy,CKT_weyl_2018_ruiyu,CKT_2018_zhaoerhai,CKT_2019_weyl_exp,CKT_2019_Chern_prl,CKT_2019_Chern_prb}, high-order topological states\cite{CKT_TI_2019_NP_corner,CKT_2018_ezawa_corner,CKT_2019_zhangxiangdong_prb,PhysRevB.99.020304,PhysRevApplied.13.064031}, and high-dimensional topological states\cite{CKT_2019_4DTI_yurui,CKT_2020_4D_yidongchong}. In most of these studies, networks with capacitors and inductors are used to obtain a particular topological state protected by crystalline symmetry or time-reversal symmetry. However, the implementation of the spin-orbit interaction (SOI) and the gauge field (including the non-Abelian gauge field) in such systems has so far received limited attention.

In this Article, we report a controllable approach to study the non-Abelian gauge field in circuit systems. We first provide circuit modules to implement non-Abelian tunnelling in the form of Pauli matrices. Then, using these building blocks, we create three representative physical systems. First, the SOI, which is time-reversal invariant. Second, the topological Chern state, which breaks time-reversal symmetry. Our building blocks can be used to construct arbitrary forms of local SU(2) gauge fields. Thus, third, we design non-reciprocal circuits and use them to achieve the real-space non-Abelian Aharonov-Bohm effect.

                                                                         
\subsection*{Building blocks of the non-Abelian gauge field}

\begin{figure}
\centering \includegraphics[width=0.99\columnwidth]{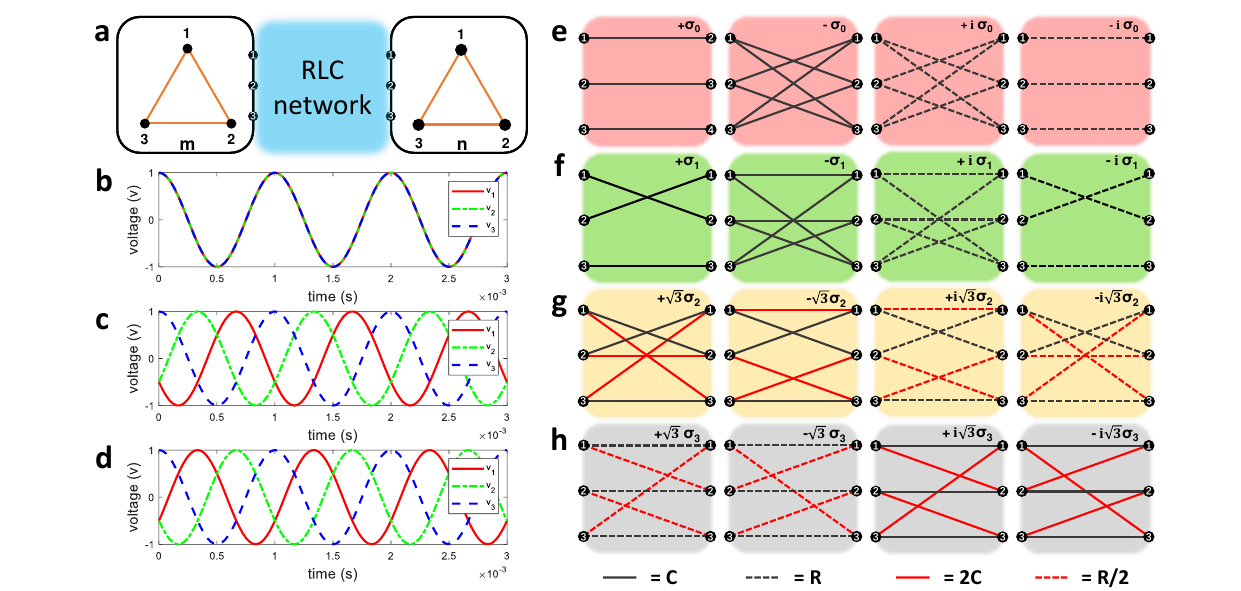}
\caption{\label{fig:fig1}\textbf{Building blocks for non-Abelian gauge field in the circuit.} 
\textbf{a}, The schematic of the pseudospin modules at cell-m and cell-n, and the connection module consisting of resistors, inductors, and capacitors. In the pseudospin module, inductors or capacitors are connected head-to-tail to form a loop structure with $C_{3}$ symmetry. 
The connection module's left and right nodes are connected to the left and right pseudospin modules, respectively. 
\textbf{b-d}, The node voltage waveforms in the form of the basis functions of the $C_{3}$ group.
\textbf{b} is the basis function $\boldsymbol{\phi}_0$ of the constant representation, \textbf{c} and \textbf{d} are the basis functions  $\boldsymbol{\phi}_{s_1}$ and $\boldsymbol{\phi}_{s_2}$ in the pseudospin space. We take the signal's frequency to be 1 kHz in the plot.
\textbf{e-h}, The list of designed connection modules $m_{\pm(i)\sigma_{0,1,2,3}}$ that give $\pm(i)\sigma_{0,1,2,3}$ types of tunneling matrices in the pseudospin space, where solid lines indicate capacitors, dashed lines indicate resistors.}
\end{figure}

We start with the Yang-Mills Hamiltonian with non-Abelian gauge field $H_{YM}=\frac{1}{2m}[(p_{x}+A_{x})^{2}+(p_{y}+A_{y})^{2}]$, where $m$ is the effective mass of the carrier and the gauge field component $A_{x,y}$ are Hermitian matrices\cite{YM_1954}. 
The vector potentials can take the form of $A_{x}=\sum_{i=0}^{3}\alpha_{i}\sigma_{i}$ and $A_{y}=\sum_{i=0}^{3}\beta_{i}\sigma_{i}$, where $\sigma_{0}$ is identity matrix and $\sigma_{1,2,3}$ are the Pauli matrices acting on a twofold degenerate space, $\alpha_{i}$ and $\beta_{i}$ are coefficients. 
Therefore, the vector potentials are not commuted with each other.
In the following, we will present a scheme to construct the twofold degenerate space and the vector potentials with the Pauli matrix type using basic electrical components.
As shown in Fig.~\ref{fig:fig1}a, we consider three identical components (capacitors or inductors) connected head to tail to form a triangle in both cell-m and cell-n.
This configuration possesses $C_{3}$ rotational symmetry. Thus the voltages at the nodes can be expanded as $\boldsymbol{{v}}=v_{0}\boldsymbol{\phi}_{0}+v_{s_{1}}\boldsymbol{\phi}_{s_{1}}+v_{s_{2}}\boldsymbol{\phi}_{s_{2}}$, where 
$\boldsymbol{\phi}_{0}=(1,1,1)^{\rm T}/\sqrt{3}$,
$\boldsymbol{\phi}_{s_1}=(\epsilon,\epsilon^*,1)^{\rm T}/\sqrt{3}$,
$\boldsymbol{\phi}_{s_2}=(\epsilon^*,\epsilon,1)^{\rm T}/\sqrt{3}$ are the basis functions of the irreducible representation of the $C_{3}$ group, and $\epsilon=e^{i2\pi/3}$ (see Supplementary Tab.~S1). 
In Fig.~\ref{fig:fig1}b-d, we present the voltage waveforms of the basis functions on nodes 1 to 3 in cell-m.
With time-reversal symmetry, we choose $\boldsymbol{\phi}_{s_1}$ and  $\boldsymbol{\phi}_{s_2}$, which are the complex conjugates of each other, to span the twofold degenerate pseudospin subspace.
In this article, cell-m and cell-n are referred to as pseudospin modules.

The designed connection modules that provide matrix form vector potentials between pseudospin modules are shown in Fig.~\ref{fig:fig1}e-h.
Considering that the typical values of the parameters of capacitance, inductance, and resistance are positive real numbers, we design a complete set of tunnelling modules that provide vector potentials in the form of $\pm\sigma_{0,1,2,3}$ and $\pm i\sigma_{0,1,2,3}$.
These modules enable the implementation of all forms of non-Abelian vector potentials for the two-band models.
More information on the modules in Fig.~\ref{fig:fig1} is given in Supplementary Section 1.


\subsection*{SOI in circuit}

The vector potentials in $H_{YM}$ can take the form of 
$A_{x}=(\alpha^{\prime}+\beta^{\prime})\sigma_{1}$ and
$A_{y}=(\alpha^{\prime}-\beta^{\prime})\sigma_{2}$. 
Substituting $A_{x,y}$ into $H_{YM}$, and applying spin rotation $\sigma_{1}\rightarrow-\sigma_{2}$ and $\sigma_{2}\rightarrow\sigma_{1}$, we get the Rashba and Dresselhaus SOI Hamiltonian
$H=\frac{p^{2}}{2m}-\alpha(p_{x}\sigma_{2}-p_{y}\sigma_{1})-\beta(p_{x}\sigma_{2}+p_{y}\sigma_{1})+const$, where $p^{2}=p_{x}^{2}+p_{y}^{2}$, and $\sigma_{1,2}$ are the Pauli matrices acting on the pseudospin space.
$\alpha=\alpha^{\prime}/m$ and $\beta=\beta^{\prime}/m$ are the Rashba and Dresselhaus SOI constants, respectively\cite{Rashba,Dresselhaus}. 
For 2D lattice systems, the Rashba and Dresselhaus SOI Hamiltonian can be written as
\begin{equation}
H_{2D-SOI}(\bm{k})=t_{0}(\cos k_{x}+\cos k_{y})-(\alpha+\beta)\sin k_{x}\sigma_{2}+(\alpha-\beta)\sin k_{y}\sigma_{1}.\label{eq:HRD}
\end{equation}
For 1D lattice systems, we get
\begin{equation}
H_{1D-SOI}(k_x)=t_{0}\cos k_{x}-(\alpha+\beta)\sin k_{x}\sigma_{2}.\label{eq:H1DSOC}
\end{equation}
Writing equations (\ref{eq:HRD}) and (\ref{eq:H1DSOC})  in real space, one can obtain hopping terms proportional to ($\sigma_{0}$, $\pm i\sigma_{2}$) and ($\sigma_{0}$,
$\pm i\sigma_{1}$) in x- and y-directions, respectively.

\begin{figure}
\centering \includegraphics[width=0.99\columnwidth]{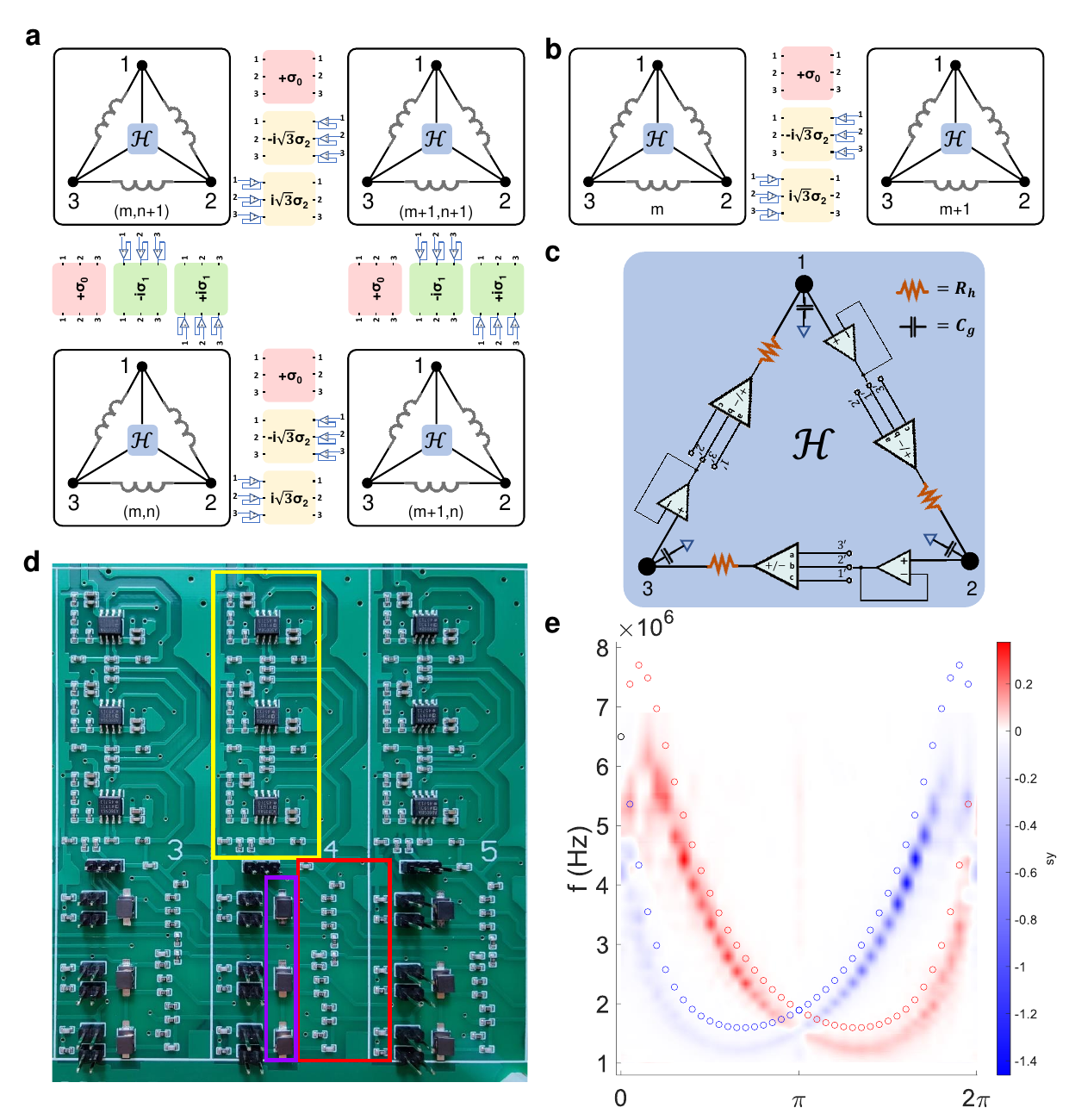}
\caption{\label{fig:fig2}\textbf{The spin-orbit interaction circuit.}
\textbf{a}, Schematic and circuit diagram of the 2D-SOI. The red, green and yellow blocks are the connection modules. The nodes of the connection modules are connected to the nodes of the neighbouring pseudo-spin modules.
The details of the connection modules are given in Fig.~\ref{fig:fig1}e-h.
The $\mathcal{H}$-modules in the pseudo-spin modules are used to eliminate the non-Hermitian terms caused by the resistors in modules $m_{\pm i\sigma_{1,2}}$ and bring the system back to Hermitian.
\textbf{b}, Schematic and circuit diagram of the 1D-SOI circuit. 
\textbf{c}, Circuit diagram of the $\mathcal{H}$-module, which consists of voltage followers (operational amplifier buffers), operational amplifier adder-subtractors, capacitors, and resistors.
\textbf{d}, A unit cell of the 1D-SOI PCB, where the yellow box indicates the $\mathcal{H}$-module, the red box indicates the $m_{\sigma_0}$ and $m_{\pm i \sigma_2}$ modules. The purple box indicates the inductors.
\textbf{e}, Experimentally measured eigenfrequency dispersions of the 1D-SOI circuit.
The red and blue colours represent the positive and negative components of the spin in the y-direction, respectively. The blocks correspond to the experimental data, and the circles represent the theoretical results.
}
\end{figure}

We now present a scheme to implement 1D- and 2D-SOI in the circuit using the building blocks given in Fig.~\ref{fig:fig1}e-h.
We choose modules $m_{\sigma_{0}}$ and $m_{\pm i\sigma_{1,2}}$ to realise the hopping terms and use the voltage followers (operational amplifier buffers) to control the hopping directions.
As the resistors used in modules $m_{\pm i\sigma_{1,2}}$ cause energy loss and yield the non-Hermitian terms, we use a sub-circuit denoted as $\mathcal{H}$ to compensate for this energy loss and bring the system back to Hermitian.
Based on these considerations, we construct the SOI circuits as shown in Fig.~\ref{fig:fig2}a-c. Kirchhoff's equations of the SOI circuit are given as
\begin{equation}
(h_{1}(\boldsymbol{k})\oplus H_{SOI}^{circuit}(\boldsymbol{k}))\tilde{\boldsymbol{v}}=\omega^{-2}(0\oplus I_{2})\tilde{\boldsymbol{v}},\label{eq:H_soc3}
\end{equation}
where $\oplus$ stands for direct sum of the constant representation space and the pseudospin space of $C_{3}$ group, $\tilde{\boldsymbol{v}}=(\tilde{v}_{1},\tilde{v}_{2},\tilde{v}_{3})^{\rm T}$ are the node voltages in the basis of the eigenfunctions of $C_{3}$ group.
$H_{SOI}^{circuit}(\boldsymbol{k})$ is the Hamiltonian in the pseudospin space.
For the 1D-SOI circuit, $H_{1D-SOI}^{circuit}({k_x})=\sum_{i=0,2}f_{i}({k_x})\sigma_{i}$, where $f_{0}({k_x})=2LC_{0}(1-\cos k_{x})/3+LC_{g}/3$, and $f_{2}({k_x})=-2L\sin k_{x}/\sqrt{3}\omega R_{x}$.
For the 2D-SOI circuit, 
$H_{2D-SOI}^{circuit}(\boldsymbol{k})=\sum_{i=0}^{2}f_{i}(\boldsymbol{k})\sigma_{i}$, where
$f_{0}(\boldsymbol{k})=2LC_{0}(2-\cos k_{x}-\cos k_{y})/3+LC_{g}/3$,
$f_{1}(\boldsymbol{k})=-2L\sin k_{y}/3\omega R_{y}$, and
$f_{2}(\boldsymbol{k})=-2L\sin k_{x}/\sqrt{3}\omega R_{x}$.
$R_{x,y}$, $C_{0,g}$, $L$ are parameters of the components.
For linear circuits, the non-Hermitian terms introduced by modules $m_{\pm i\sigma_{1,2}}$ are linear.
Therefore, in the $\mathcal{H}$-module, we can use addition, subtraction, and multiplication functions of the operational amplifier to precisely compensate for these non-Hermitian terms and guarantee the Hermitian form of the Hamiltonian $H_{1D-SOI}^{circuit}({k_x})$ and $H_{2D-SOI}^{circuit}({\bm k})$.

To verify our theoretical designs, we fabricate a 1D-SOI printed circuit board (PCB) as shown in Fig.~\ref{fig:fig2}d and measure its eigenfrequency dispersions.
As shown in Fig.~\ref{fig:fig2}e, the frequency bands indicate the SOI type of splitting at $k_x=0$ and $k_x=\pi$ points, which is consistent with the theoretical calculations. Details of the experiment are presented in the methods section.
The details of the 1D-SOI and 2D-SOI circuits, the proof of the stability of the circuits, the derivation of the equation (\ref{eq:H_soc3}), and the spin information of the eigenstates are provided in Supplementary Section 2.
Our scheme presented above can be easily generalised to lattice models in three or higher spatial dimensions to study various novel physics related to SOI.

\subsection*{Topological Chern state in circuit}

We now discuss the Chern insulator's design scheme in a circuit that breaks the time-reversal symmetry. We start with a Chern insulator Hamiltonian 
\begin{equation}
H_{Chern}(\bm{k})=\sum_{i=1}^{3}d_{i}(\bm{k})\sigma_{i},\label{eq:H_Chern}
\end{equation}
where 
$d_{1}(\bm{k})=t_{1}\cos k_{y}$, 
$d_{2}(\bm{k})=t_{2}\cos k_{x}$,
$d_{3}(\bm{k})=m_{0}+t_{3}(\sin k_{x}+\sin k_{y})$. $t_{1,2,3}$
are the hopping parameters, and $m_{0}$ is the mass term. The Hamiltonian has non-zero Chern number if $0<\lvert m_{0}\rvert <2\lvert t_{3}\rvert$.
The Chern model can be explored with a square lattice described by the tight-binding Hamiltonian
\begin{equation}
H_{Chern}=\sum_{m,n}(c_{m+1,n}^{+}\hat{U}_{x}c_{m,n}+c_{m,n+1}^{+}\hat{U}_{y}c_{m,n}+c_{m,n}^{+}\hat{M}c_{m,n}+h.c.),
\end{equation}
where $\hat{U}_{x}=t_{2}/2\sigma_{2}-it_{3}/2\sigma_{3}$ and $\hat{U}_{y}=t_{1}/2\sigma_{1}-it_{3}/2\sigma_{3}$ are non-Abelian hopping operators. $\hat{M}=m_{0}\sigma_{3}$ is the mass term and the two-component operator $c_{m,n}^{+}$ creates a particle at site (m,n) in the pseudospin space.

\begin{figure}
\centering \includegraphics[width=0.99\columnwidth]{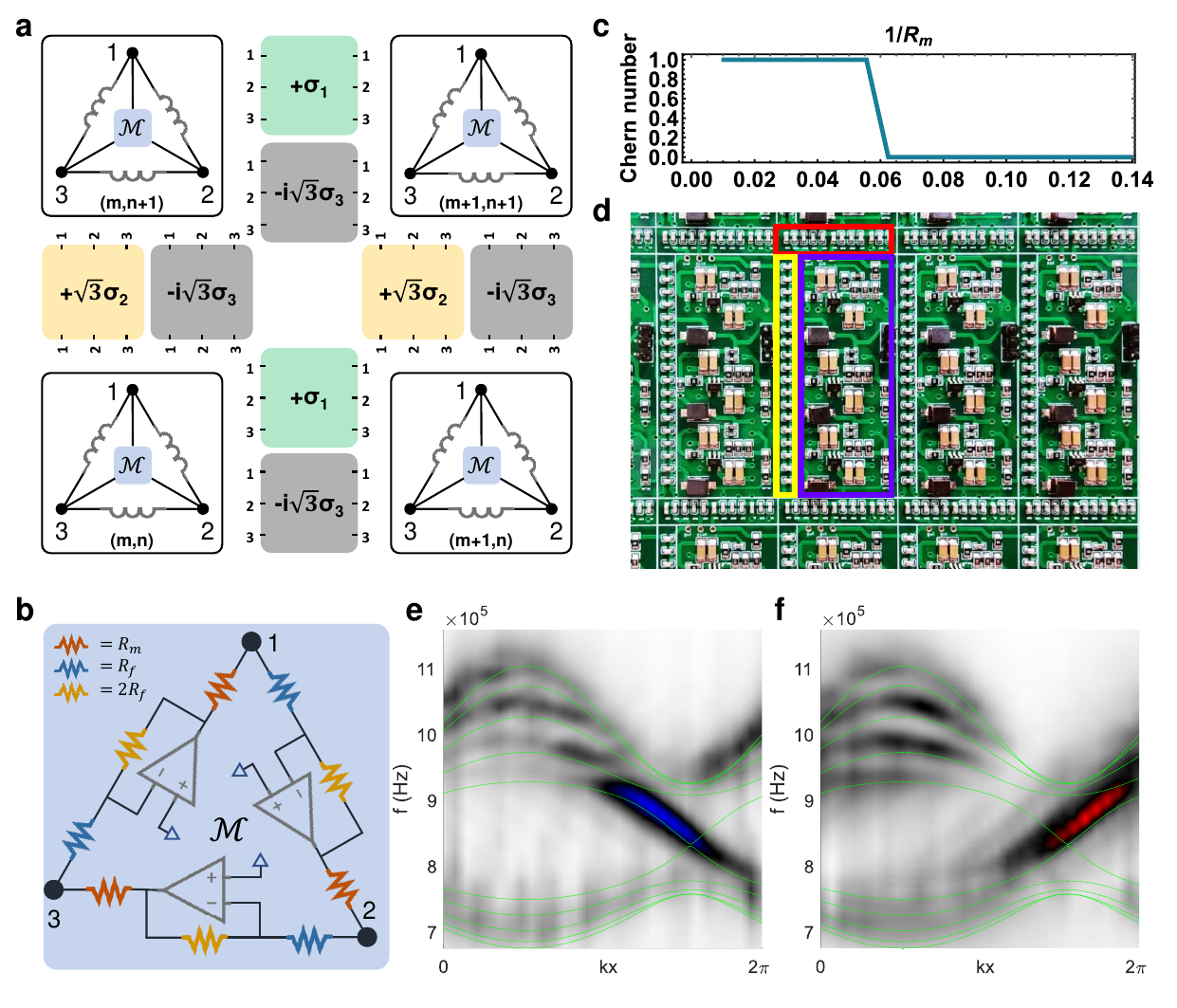}
\caption{\label{fig:fig3}\textbf{Topological Chern circuit.} 
\textbf{a}, Schematic and circuit diagram of the Chern circuit. The yellow, green and grey blocks are the connection modules. The nodes of the connection modules are connected to the nodes of the neighbouring pseudo-spin modules. The $\mathcal{M}$-modules in the pseudo-spin modules are used to generate topological mass terms and break time-reversal symmetry.
\textbf{b}, Details of the $\mathcal{M}$-module, which consists of inverting operational amplifier and resistors  $R_{m}$.
\textbf{c}, The phase diagram of the Chern number as a function of $R_{m}$.
\textbf{d}, Printed circuit board layout of the topological Chern circuit. 
The purple box shows the inductors and mass term module. 
The yellow (red) box indicates the $m_{\sigma_{1}}+m_{-i\sigma_{3}}$ ($m_{\sigma_{2}}+m_{-i\sigma_{3}}$) module. 
\textbf{e}, Comparison of the experimentally measured edge states, excited by a source at $y=1$ boundary, with the theoretical results (green dashed curves) for a 30$\times$5 unit cells system. 
The main component of the spin of the boundary state is along the y-direction, where red indicates $s_y>0$ and blue indicates $s_y<0$. 
\textbf{f}, Experimentally measured edge states excited by a source at $y=5$ boundary.}
\end{figure}

In the Chern circuit, the hopping terms are implemented with modules $m_{\sigma_{1,2}}$ and $m_{-i\sigma_{3}}$ as illustrated in Fig.~\ref{fig:fig3}a, and the mass term is realized by using the inverting operational amplifier as shown in Fig.~\ref{fig:fig3}b.
Kirchhoff's equations of the Chern circuit are given as 
\begin{equation}
(h_{1}(\bm{k})\oplus H_{Chern}^{circuit}(\bm{k}))\tilde{\bm v}=\omega^{-2}(0\oplus I_{2})\tilde{\bm v},\label{eq:H_Chern_CKT}
\end{equation}
where $H_{Chern}^{circuit}(\bm{k})=\sum_{i=0}^{3}g_{i}(\bm{k})\sigma_{i}$, with 
$g_{0}(\bm{k})=L(C_{m}+2C_{1}/3+2C_{2}+4C_{3})$, 
$g_{1}(\bm{k})=-2LC_{1}\cos k_{y}/3$,
$g_{2}(\bm{k})=-2LC_{2}\cos k_{x}/\sqrt{3}$, and 
$g_{3}(\bm{k})=-L/\sqrt{3}\omega R_{m}-2LC_{3}(\sin k_{x}+\sin k_{y})/\sqrt{3}$.
$C_{1,2,3,m}$, $R_{m}$ and $L$ are parameters of the components.
The mass term $-L/\sqrt{3}\omega R_{m}$ in $g_{3}(\bm{k})$ is contributed by the $\mathcal{M}$-module, which breaks the time-reversal symmetry and opens the topologically non-trivial gap.
Hamiltonian $H_{Chern}$ in equation (\ref{eq:H_Chern}) and $H_{Chern}^{circuit}$ in equation (\ref{eq:H_Chern_CKT}) have the same mathematical structure except for the mass term, where the mass term
in $H_{Chern}^{circuit}$ is related to eigenfrequency $\omega$.
To identify the topological properties of the circuit system, we use Green's function method to calculate the Chern number. The obtained phase diagram for the Chern number as a function of $R_{m}$ is plotted in Fig.~\ref{fig:fig3}c.

We can examine the topological nature of the circuit by measuring the chiral edge states at the system's boundary, which is related to the Chern number of the bulk system. 
We fabricate the Chern PCB containing $30\times5$ unit cells with periodic boundary conditions in the x-direction and open boundary conditions in the y-direction.
The circuit structure in a unit cell is shown in Fig.~\ref{fig:fig3}d.
The frequency dispersion is shown in Fig.~\ref{fig:fig3}e-f, where the black blocks are experimental results, and the green dashed curves are theoretical results. The red and blue colours indicate the components of the spin in the $\pm y$ direction on the boundary state.
It is clear that the frequency band structure has a bulk gap and exhibits chiral propagating edge modes that traverse the gap. 
The experimental data show that we can only detect the edge states located on the $y=1$ ($y=5$) boundary with the excitation source applied at the same boundary. In contrast, the edge states on the opposite boundary can not be excited. Details on the experiment can be found in the methods section.
Due to parasitic effects in electronic components, such as the internal resistance of inductors and connection wires, there is an inevitable broadening in the measured bands. 
This issue can be improved by choosing inductors with lower internal resistance.
The details of the Chern circuit, the proof of its stability, the derivation of the equation (\ref{eq:H_Chern_CKT}), and the calculation of the Chern number can be found in Supplementary Section 3.
In the Supplementary information, we also provide a scheme to realise a Chern circuit with positive topological mass term using operational amplifiers with integrator form.


\subsection*{The non-reciprocal circuit and the real space non-Abelian Aharonov-Bohm effect}
Using the building blocks given in Fig.~\ref{fig:fig1}, we have implemented SOI and Chern insulators in the circuit system, which are novel phenomena generated by non-Abelian gauge field in momentum space. 
However, an essential feature of the non-Abelian gauge field that the phase of the wave function is related to the order of the gauge fields that the wave passes through is not directly manifested in the above two phenomena.
To explicitly characterise this property, we design non-reciprocal circuits and show that they tune the phase of the signals in a non-Abelian manner in the real space.

\begin{figure}
\centering \includegraphics[width=0.99\columnwidth]{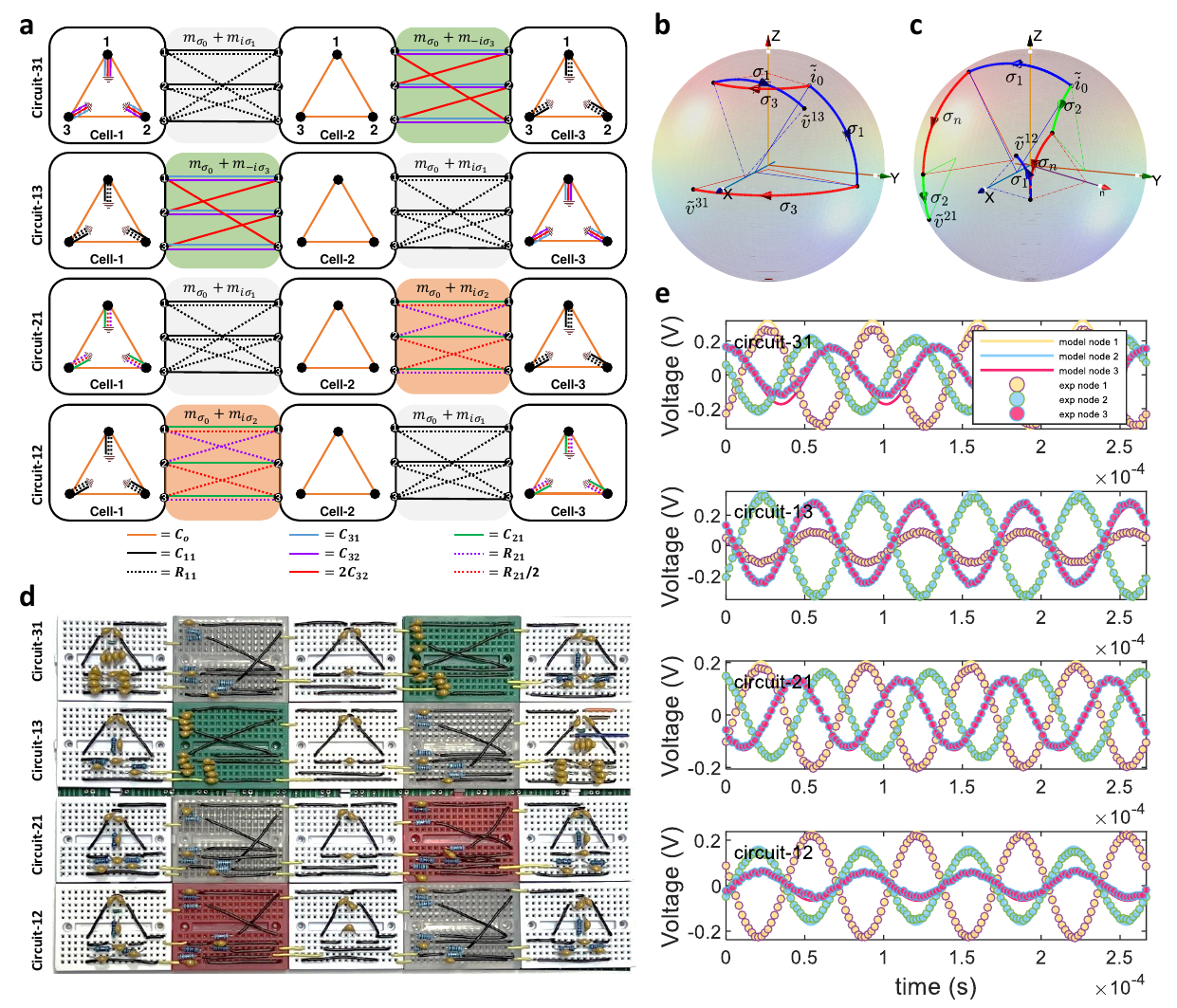}
\caption{\label{fig:fig4}\textbf{The non-reciprocal circuit that generates non-Abelian phase factor.}
\textbf{a}, Schematic diagram of circuit-31, -13, -21 and -12.
The capacitors in cell-1, -2 and -3 form a triangle structure.
The nodes in cell-1 and cell-3 are grounded with capacitors and resistors to make the three cells have the same on-site frequency.
The gray module consists of  $m_{\sigma_{0}}$ and $m_{i\sigma_{1}}$.
The green module consists of $m_{\sigma_{0}}$ and $m_{-i\sigma_{3}}$.
The red module consists of $m_{\sigma_{0}}$ and $m_{i\sigma_{2}}$.
\textbf{b-c}, The same initial
state $\tilde{\bm i}_{0}$ leads to different final states under the non-Abelian gauge field. 
\textbf{b} for circuit-31 and circuit-13 and \textbf{c} for circuit-21 and circuit-12.
\textbf{d}, The constructed circuit-31, -13, -21 and -12 in breadboards.
\textbf{e}, Output voltages at the three nodes of cell-3 with the same input at cell-1, where the dots indicate experimental data and the lines indicate theoretical results.
The component parameters and experimental details are provided in the methods section.}
\end{figure}

We consider four circuits consisting of cells-1, -2, -3 and two connection modules for each as shown in Fig.~\ref{fig:fig4}a.
In circuit-31, cell-1 and cell-2 are connected by modules $m_{\sigma_{0}}$ and $m_{i\sigma_{1}}$. In the spin subspace, the connection between cell-1 and cell-2 can be written in the form of $m_{01}=i(\omega C_{11}\sigma_{0}+R_{11}^{-1}i\sigma_{1})=i\alpha e^{i\theta_{1}\sigma_{1}}$, where $\alpha=\sqrt{\omega^{2}C_{11}^{2}+1/R_{11}^{2}}$ and $\theta_{1}=\arctan(1/\omega C_{11}R_{11})$.
The connection modules between cell-2 and cell-3 are $m_{\sigma_{0}}$ and $m_{-i\sigma_{3}}$, which give a connection in the form of $m_{03}=i(\omega C_{31}\sigma_{0}-\sqrt{3}\omega C_{32}i\sigma_{3})=i\beta e^{i\theta_{3}\sigma_{3}}$, where $\beta=\sqrt{\omega^{2}C_{31}^{2}+3\omega^{2}C_{32}^{2}}$ and $\theta_{3}=\arctan(-\sqrt{3}C_{32}/C_{31})$. 
$C_{11}$, $R_{11}$, $C_{31},$ and $C_{32}$ are parameters of the components in the connection modules. 
Exchanging the order of $m_{01}$ and $m_{03}$ in circuit-31 gives circuit-13. 
Since $m_{01}$ and $m_{03}$ are not commutative, circuit-31 and circuit-13 will yield different outputs for the same input signals. 
To prove this result rigorously, we consider inputting currents at cell-1 and detecting voltages at cell-3. 
The transfer impedance equation for the input and output is
\begin{equation}
\tilde{\bm v}_{out}^{31}=(z_{1}\oplus Z^{31}_{s})\tilde{\bm i}_{in}=(z_{1}\oplus z_{2}e^{-i\theta_{3}\sigma_{3}}e^{i(\theta_{1}-\phi)\sigma_{1}})\tilde{\bm i}_{in}\label{eq:-23}
\end{equation}
for circuit-31 and
\begin{equation}
\tilde{\bm v}_{out}^{13}=(z_{1}\oplus Z^{13}_{s})\tilde{\bm i}_{in}=(z_{1}\oplus z_{2}e^{i(\theta_{1}-\phi)\sigma_{1}}e^{-i\theta_{3}\sigma_{3}})\tilde{\bm i}_{in}\label{eq:-24}
\end{equation}
for circuit-13, where $z_{1}=\lambda_{1}^{-1}/2$, $z_{2}=-\alpha\beta/\gamma$,
$\tan\phi=\gamma_{1}/\gamma_{0}$, $\gamma=\sqrt{\gamma_{0}^{2}+\gamma_{1}^{2}}$,
$\gamma_{0}=\lambda_{2}(\beta^{2}-\omega^{2}C_{11}^{2}+1/R_{11}^{2}-\lambda_{2}^{2})$,
$\gamma_{1}=-2\omega C_{11}/R_{11}$, $\lambda_{1}=-i\omega(C_{11}+C_{31}+3C_{32})-2/R_{11}$
and $\lambda_{2}=\lambda_{1}-i\omega3C_{o}$.
$Z^{13(31)}_s$  denotes the transfer impedance matrix in the pseudospin space. Equations (\ref{eq:-23}) and (\ref{eq:-24}) show that the pseudospin component of the signal is confined to this degenerate pseudospin space as it flows through the circuit and acquires a phase modulation in the form of a non-Abelian gauge field.
For the same initial state $\tilde{\bm i}_{0}$, passing through circuit-31 and circuit-13, could yield different final states $\tilde{\bm v}^{13}$ and $\tilde{\bm v}^{31}$, as shown in Fig.~\ref{fig:fig4}b.

In circuit-21, the connection modules are $m_{\sigma_{0}}+m_{i\sigma_{1}}$ between cell-1 and cell-2, and $m_{\sigma_{0}}+m_{i\sigma_{2}}$ between cell-2 and cell-3. 
The order of the connection modules is exchanged in circuit-12 as shown in Fig.~\ref{fig:fig4}a. 
The transfer impedance equations are obtained as 
\begin{equation}
\tilde{\bm v}_{out}^{21}=(z_{3}\oplus Z^{21}_{s})\tilde{\bm i}_{in}=(z_{3}\oplus z_{4}e^{i\theta_{2}\sigma_{2}}e^{-i\phi(\hat{\bm n}\cdot\vec{\sigma})}e^{i\theta_{1}\sigma_{1}})\tilde{\bm i}_{in}\label{eq:-40}
\end{equation}
for circuit-21 and 
\begin{equation}
\tilde{\bm v}_{out}^{12}=(z_{3}\oplus Z^{12}_{s})\tilde{\bm i}_{in}=(z_{3}\oplus z_{4}e^{i\theta_{1}\sigma_{1}}e^{-i\phi(\hat{\bm n}\cdot\vec{\sigma})}e^{i\theta_{2}\sigma_{2}})\tilde{\bm i}_{in}\label{eq:-41}
\end{equation}
for circuit-12, where $\phi$ and $\hat{\bm n}$ are functions of $\theta_{1}$
and $\theta_{2}$. 
The processes for the same initial state modulating by the circuit-21 and circuit-12 yielding different final states are presented in Fig.~\ref{fig:fig4}c, which indicates that circuit-21 and circuit-12 also perform a non-Abelian type phase modulation to the input signal.
The specific expressions for $\theta_{1}$, $\theta_{2}$,
$\phi$, $\hat{\bm n}$, $z_{1,2,3,4}$ and the derivation details of equations (\ref{eq:-23}-\ref{eq:-41}) are given in Supplementary Section 4.

To check the above theoretical results, we prepare these four circuits on the breadboard as shown in Fig.~\ref{fig:fig4}d. 
The AC currents are inputted at the three nodes of cell-1, and the output voltages are measured at the nodes in cell-3.
As shown in Fig.~\ref{fig:fig4}e, the experimental curves and the theoretical ones are in good agreement, where the DC components of the experimental results are deducted. 
Details of the four circuits, the experimental measurements and the value of the DC components for each experiment are given in the methods section.

\begin{figure}
\centering \includegraphics[width=0.99\columnwidth]{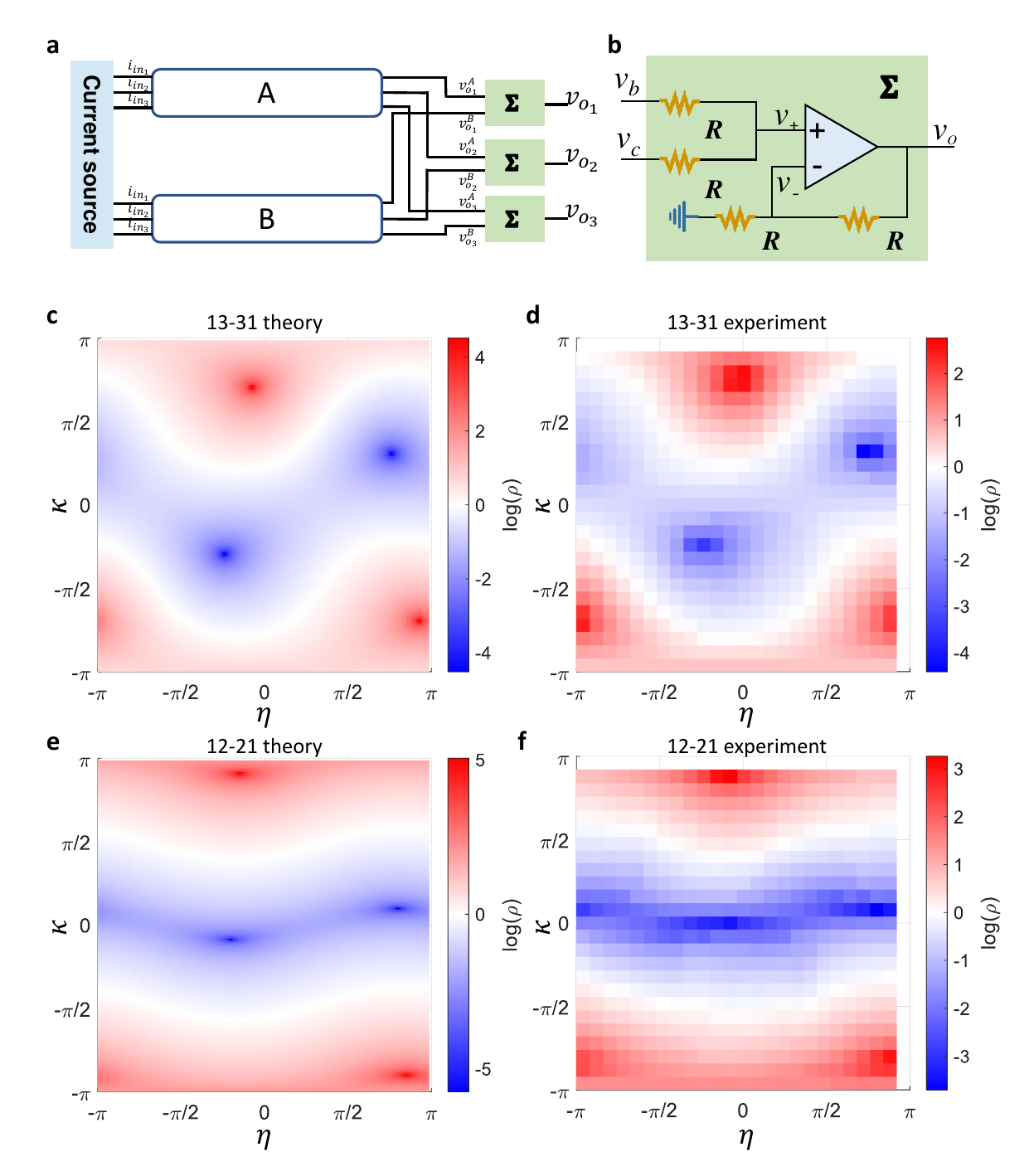}
\caption{\label{fig:fig5}
\textbf{The real space non-Abelian Aharonov-Bohm effect.} 
\textbf{a}, Schematic diagram of the circuit for the real space non-Abelian Aharonov-Bohm effect. Modules A and B are circuits 13 and 31 or 12 and 21 as designed in Fig.~\ref{fig:fig4}. The $\Sigma$-module is an operational amplifier voltage adder.
\textbf{b}, Details of the $\Sigma$-module, where the input-output relation is given as $v_{o}=v_{b}+v_{c}$.
\textbf{c}, \textbf{d} (\textbf{e}, \textbf{f}), The theoretical and experimental results of the contrast functions $\rho$ with the A-B modules are composed of circuits 13-31 (12-21).}
\end{figure}

The phase factors in equations (\ref{eq:-23}-\ref{eq:-41}) are the keys for realising the non-Abelian Aharonov-Bohm effect.
In Fig.~\ref{fig:fig5}a, we design the circuit system to implement the non-Abelian Aharonov-Bohm effect in real space.
The current source provides the input currents with controllable phase and amplitude.
The $\Sigma$-module is an operational amplifier voltage adder detailed in Fig.~\ref{fig:fig5}b. 
Modules A and B are circuits 13 and 31 or circuits 12 and 21. 
The signals ${\boldsymbol{i}}_{in}$ flowing through modules A and B undergo phase modulation as stated in equations (\ref{eq:-23}-\ref{eq:-41}) and then are collected by the $\Sigma$-modules to produce the final output voltages ${\boldsymbol{v}}_{o}$.
In the pseudospin space, the entire process described above can be expressed with the following equation
\begin{equation}
\tilde{\boldsymbol{v}}_{o}=((z_{1}+z_{3})\oplus(Z_{s}^{A}+Z_{s}^{B}))\tilde{\boldsymbol{i}}_{in}.
\end{equation}
We define the intensity contrast function
$\rho= \lvert\boldsymbol{v}_{o_{s_1}}\rvert/ \lvert\boldsymbol{v}_{o_{s_2}}\rvert$ 
to examine the interference of the output signals, where 
$\boldsymbol{v}_{o_{s_1}}$ ($\boldsymbol{v}_{o_{s_2}}$) 
is the spin-up (down) component of the output voltages.
For modules A-B taking circuits 13-31, we get
\begin{equation}
\boldsymbol{v}_{o_{s_1}}=\left(\cos\frac{\eta}{2}e^{-i(\theta_3+\kappa)}\cos(\theta_1-\phi)+i\sin\frac{\eta}{2}\cos\theta_3\sin(\theta_1-\phi)\right)e^{+i\frac{\kappa}{2}},
\end{equation}
and
\begin{equation}
\boldsymbol{v}_{o_{s_2}}=\left(\sin\frac{\eta}{2}e^{+i(\theta_3+\kappa)}\cos(\theta_1-\phi)+i\cos\frac{\eta}{2}\cos\theta_3\sin(\theta_1-\phi)\right)e^{-i\frac{\kappa}{2}},
\end{equation}
where the orientation of the spin in the input currents is characterized by angles $\eta$ and $\kappa$ on the Bloch sphere as $\tilde{\boldsymbol{i}}_{in}=(i_{0},i_{s}\cos\eta/2,i_{s}e^{i\kappa}\sin\eta/2)$.
The intensity contrast $\rho$ with respect to $\eta$ and $\kappa$ are shown in Fig.~\ref{fig:fig5}c-f, where the theoretical and experimental results are in good agreement.
The expressions of $\boldsymbol{v}_{o_{s_{1,2}}}$ in the case of chosen circuits 12-21 and more details about the non-Abelian Aharonov-Bohm circuit are presented in Supplementary Section 4.



\subsection*{Conclusions}
We have reported a scheme to implement non-Abelian gauge fields in circuit systems, which we use to create the SOI, topological state, and real space non-Abelian Aharonov-Bohm effect. 
In the alternative Chern circuit scheme\cite{CKT_2018_zhaoerhai}, which involves only inductors and capacitors, at least six nodes in one unit cell are required to create the complex hoppings for a two-band Chern insulator. Our scheme successfully reduces the degrees of freedom in the unit cell, at the price of introducing an operational amplifier, which increases the complexity of the circuit structure to a degree. However, this allows us to use the computational functions of the operational amplifier in combination with the designed modules to generate rich physical states in the circuit. In the SOI and Chern circuits, we showed that the operational amplifiers can be used to transform a system from non-Hermitian to Hermitian and design topological time-reversal symmetry breaking terms. Our scheme could also be used to study non-Hermitian physics\cite{NM_PT_2019,nonHermitian_NJP2021} as well as a wide variety of non-Abelian physics, including the non-Abelian Aharonov-Casher effect, the Hofstadter's moth\cite{Hofstadter_moth_2005}, and the motion of particles in non-Abelian gauge fields\cite{Xiaodi_RMP_2010}.

\section*{Methods}
\textbf{PCB preparation of the 1D-SOI circuit.}
The 1D-SOI circuit was implemented on FR4 printed circuit board according to Fig.~\ref{fig:fig2}b and c. 
The parameters of the devices are chosen as follows.
Inductors $L$ are 1.8 $\mu H$ with $\pm5\%$
tolerance and 53 $m\Omega$ series resistance. 
Resistors $R_{x}$ in modules $m_{\pm i\sigma_{2}}$ are 20 $\Omega$ with $\pm1\%$ tolerance,
Capacitors $C_{0}$ in modules $m_{\sigma_{0}}$ are 2.7 nF with $\pm5\%$ tolerance.
Capacitors $C_{g}$ are 1 nF with $\pm5\%$ tolerance.
Resistors $R_{h}$ in module $\mathcal{H}$ are realized by connecting two $R_{x}$ in parallel. 
Resistors $R_{f}$ are 806 $\Omega$ with $\pm1\%$ tolerance.
The operational amplifiers are AD8058.
The 1D-SOI PCB with 20 unit cells in the x-direction is shown in Supplementary Fig.~S3d.
The voltages at each node, including the amplitude and phase, are probed by the Rohde\&Schwarz vector network analyser ZNL6 5kHz-6GHz.

\vspace{10pt}
\noindent
\textbf{PCB preparation of the Chern circuit.}
The Chern circuit was implemented on FR4 printed circuit board according to Fig.~\ref{fig:fig3}a and b. 
Inductors $L$ are 1.8 $\mu H$ with $\pm5\%$ tolerance and 53 $m\Omega$ series resistance.
Capacitors $C_1$ in module $m_{\sigma_{1}}$ are 6.8 nF with $\pm5\%$ tolerance.
Capacitors $C_{2,3}$ in modules $m_{\sigma_{2}}$ $m_{-i\sigma_{3}}$ are 2.7 nF with $\pm5\%$ tolerance.
The resistors $R_{m}$ in  $\mathcal{M}$-module are 30 $\Omega$ with $\pm1\%$ tolerance.
Resistors $R_{f}$ are 7.87 $k\Omega$ with $\pm1\%$ tolerance.
The operational amplifiers are AD8057. 
The Chern PCB with 15-by-5 unit cells is shown in Supplementary Fig.~S4c.
The voltages at each node, including the amplitude and phase, are probed by Rohde\&Schwarz vector network analyser ZNL6 5kHz-6GHz.

\vspace{10pt}
\noindent
\textbf{Circuit board preparation and the measurement of the real space non-Abelian Aharonov-Bohm effect.}
The circuit for real space non-Abelian Aharonov-Bohm effect is implemented on a breadboard as shown in Fig.~\ref{fig:fig4}d and Fig.~\ref{fig:fig5}a. 
The capacitors
$C_{11}$,
$C_{21}$,
$C_{31}$,
$C_{32}$,
$C_{o}$ are chosen with the same parameter, where the capacitance is 2.7 nF with $\pm20\%$ tolerance. 
Resistors
$R_{11}$,
$R_{21}$ are the same, where their resistance is 1 k$\Omega$ with $\pm1\%$ tolerance. 
Due to the limited size of the breadboard, where two 1 k$\Omega$ resistors are needed in parallel, we replaced them with a resistor of 510 $\Omega$ with $\pm1\%$ tolerance. 
Where three 1 k$\Omega$ resistors are needed in parallel, we replaced them with a 330 $\Omega$ resistor with $\pm1\%$ tolerance. 
The operational amplifier is AD8047.
The current frequency is 15 kHz, which is generated by the current source instrument Keithley 6221.
The voltage is measured by Rigol oscilloscope MSO5354. To compare with the theoretical results, we deduct the DC components of the experimental data, which are -30 mV for circuit-31, -24 mV for circuit-13, -12 mV for circuit-21, and -17 mV for circuit-12. The DC errors arise for two reasons.
One is that the vertical accuracy of the oscilloscope is on the order of 10 mV. 
Another reason is that in our experiment, we use a single current source to supply current at the three input ports in three separate steps. 
Because the circuit is linear, the total output voltage is the sum of the output voltages obtained from the three measurements described above. 
However, the cost of this measurement method is that a large cumulative error is obtained. If three current sources supply the input currents simultaneously, and a more accurate oscilloscope is used, the DC components of the measurement will be reduced. 
Nevertheless, since the important information in the experimental data is the amplitude and phase of the AC components, the presence of the DC components does not affect our conclusions.

\section*{Data availability}
The data that support the findings of this study are available from the corresponding author upon reasonable request.

\section*{Code availability}
The codes that support the findings of this study are available from the corresponding author upon reasonable request.

\section*{Acknowledgements}
We thank H.M. Weng, C. Fang, D. Zhang, H. Wang, and S. Chang for valuable discussions. 
R.Y. acknowledges support from the National Key Research and Development Program of China (No. 2017YFA0304700 and No. 2017YFA0303402) and the National Natural Science Foundation of China (No. 11874048), and the Beijing National Laboratory for Condensed Matter Physics.
F.Q.S. acknowledges support from the National Natural Science Foundation of China (No. 92161201, No. 12025404, No. 11904165, No. 11904166).

\section*{Author contributions}
R.Y. supervised the project.
J.X.W., Z.W. and R.Y. drew the circuit diagram of the SOI PCB.
Z.W. and R.Y. measured the experimental data of the SOI circuit. 
Z.P.Y., Y.J.H. and R.Y. drew the circuit diagram of the Chern PCB.
Z.W., T.Y.W. Z.P.Y., Y.J.H., and Z.Y.S. measured the experimental data of the Chern circuit.
Z.W and Y.C.B., S.Z., and F.C.F. measured the experimental data of the non-Abelian Aharonov-Bohm effect. 
J.X.W. wrote part of the data acquisition program for the experimental instruments. 
F.Q.S. and R.Y. analysed the data.

\section*{Competing Interests}
The authors declare no competing interests.


\section*{Figure Legends/Captions}

\clearpage

\clearpage




\end{document}